\documentclass[11pt,preprint]{aastex}
\input psfig

\def\kmsm{\,{\rm km\, s^{-1}\, Mpc^{-1}} }

\begin{document}

\title{Measuring Supermassive Black Holes in Distant Galaxies with
Central Lensed Images}

\author{David Rusin\altaffilmark{1}, Charles R.\
Keeton\altaffilmark{2}, Joshua N.\ Winn\altaffilmark{3,4}}

\altaffiltext{1}{Department of Physics \& Astronomy, University of
  Pennsylvania, 209 So.\ 33rd St., Philadelphia, PA 19104-6396}
\altaffiltext{2}{Department of Physics \& Astronomy, Rutgers
  University, 136 Frelinghuysen Road, Piscataway, NJ 08854}
\altaffiltext{3}{Harvard-Smithsonian Center for Astrophysics, 60
  Garden St., Cambridge, MA 02138}
\altaffiltext{4}{Hubble Fellow}

\begin{abstract}

  The supermassive black hole at the center of a distant galaxy can be
  weighed, in rare but realistic cases, when the galaxy acts as a
  strong gravitational lens.  The central image that should be
  produced by the lens is either destroyed or accompanied by a second
  central image, depending on the mass of the black hole.  We
  demonstrate that when a central image pair is detected, the mass of
  the black hole can be determined with an accuracy of $\la 0.1$ dex,
  if the form of the smooth mass distribution near the galaxy core is
  known. Uncertainty in the central mass distribution introduces a
  systematic error in the black hole mass measurement. However, even
  with nearly complete ignorance of the inner mass distribution, the
  black hole mass can still be determined to within a factor of
  10. Central image pairs should be readily observable with future radio
  interferometers, allowing this technique to be used for a census of
  supermassive black holes in inactive galaxies at significant
  redshift ($0.2 \la z \la 1.0$).

\end{abstract}

\keywords{galaxies: nuclei---black hole physics---gravitational
lensing}

\section{Introduction}

Supermassive black holes (SMBHs) reside at the centers of most
galaxies, and are the power sources for active galactic nuclei
(AGNs). For nearby galaxies, empirical correlations have been
discovered between the mass of the SMBH and various galactic
properties such as the bulge mass (Laor 2001), velocity dispersion
(Ferrarese \& Merritt 2000; Gebhardt et al.\ 2000; Tremaine et al.\
2002), luminosity (Magorrian et al.\ 1998; McLure \& Dunlop 2001,
2002), and concentration (Graham et al.\ 2001). The existence of these
correlations suggests a relationship between black hole growth and
galaxy formation (e.g., Kauffmann \& Haenhelt 2000; Monaco et al.\
2000; Wyithe \& Loeb 2002; Haiman et al.\ 2004).

An important step forward would be to trace the evolution of these
correlations over cosmic time (see, e.g., Treu et al.\ 2004; Robertson
et al.\ 2005). For this, we need a way to measure central black hole
masses in galaxies at significant redshift. Two techniques have
provided most of the existing black hole mass measurements. First, in
normal galaxies, the SMBH mass can be determined from the kinematics
of stars or gas near the galaxy center (Magorrian et al.\ 1998;
Gebhardt et al.\ 2003). Although this method offers very accurate
masses, it requires spectroscopy with high spatial resolution, which
can only be achieved for nearby galaxies. Second, in reverberation
mapping of active galaxies (Blandford \& McKee 1982; Kaspi et al.\
2000; Peterson et al.\ 2004), the orbital parameters of gas clouds can
be estimated from the width of the broad emission lines and the time
lag between continuum and line variability, yielding a measurement of
the SMBH mass. This method works at higher redshift, but it cannot be
applied to normal, quiescent galaxies.

Gravitational lensing supplements these techniques by offering a way
to detect and measure SMBHs in ordinary galaxies at intermediate
redshift ($0.2 \la z \la 1.0$). The starting point is the generic
prediction that the multiple lensed images of a background object
should include a faint image that appears close to the center of the
lens (Burke 1981). A SMBH in the lens galaxy will alter the properties
of this image, in one of two ways (Mao et al.\ 2001). For certain SMBH
masses and source positions, the black hole destroys the central
image. If that does not happen, then the SMBH creates an {\em
additional} image.

Radio observations seem to be the likeliest route to finding central
images, because of the necessary angular resolution and dynamic range,
and because dust extinction is not a problem at radio
wavelengths. Bowman et al.\ (2004) recently calculated the probability
of finding central image pairs produced by a SMBH, and concluded that
the next generation of radio telescopes is needed to discover them in
large numbers. In this Letter, we investigate what will be learned
when a central image pair is detected: namely, with what accuracy can
the mass of the SMBH be measured?  The issue is timely because the
best candidate for a central image has recently been found in the lens
system PMN~J1632$-$0033 (Winn et al.\ 2003b, 2004). Since the central
image has apparently not been destroyed in that system, there must be
a faint companion image if the lens galaxy hosts a SMBH.

Section 2 reviews the phenomenology of central images. Section 3
presents an analysis of simulated central image systems, and \S~4
summarizes and discusses these results. Standard values for the
cosmological parameters ($\Omega_m = 0.3$, $\Omega_{\Lambda} = 0.7$,
and $H_0 = 65 \kmsm$) are assumed for all calculations.

\section{The Phenomenology of Central Images}

In the language of lensing theory (see, e.g., Kochanek et al.\ 2004),
the formation of a central image requires the existence of a radial
critical curve. For spherical models, this is equivalent to a solution
of the equation $d\alpha(R)/dR=1$, where $\alpha(R)$ is the deflection
angle and $R$ is the image-plane radius. Many commonly used mass
models have exactly one solution, such as the nonsingular isothermal
sphere (NIS), and single or broken power-law profiles with an inner
slope that is shallower than isothermal ($\rho \propto r^{-\beta}$
with $\beta<2$). These models produce one central image, and have been
described in detail by Wallington \& Narayan (1993), Rusin \& Ma
(2001) and Keeton (2003).

As an example, Figure~1 shows the deflection profile of the NIS model,
which produces three images of any source within the radial caustic.
We denote these images as A (minimum), B (saddle point) and C
(maximum).\footnote{The terminology indicates whether the image forms
at a minimum, saddle point, or maximum in the time delay surface
(Schneider 1985, Blandford \& Narayan 1986).} The addition of a
compact mass at the lens center modifies the properties of central
images because it causes the deflection angle to diverge as $R
\rightarrow 0$, creating a second radial critical curve. Consequently,
a sufficiently misaligned source ($R_{s,1}$) produces a second central
image:\ a saddle point, denoted as D. In contrast, no central images
form if the source is well aligned ($R_{s,2}$). This illustrates the
general conclusion that the SMBH either destroys the central image, or
introduces a second one (Mao et al.\ 2001).

For a given profile, detectable central image pairs only form over a
narrow range of black hole masses. We illustrate this by considering a
spherical power-law (PL, $\rho \propto r^{-\beta}$) galaxy with a SMBH
at $z=0.5$, and a source at $z=1.5$. The models are normalized to a
typical Einstein radius of $0\farcs75$. Our definition of
``detectable'' is $\mu_C / \mu_A > 2\times 10^{-3}$ and $\mu_D / \mu_C
> 10^{-2}$, where $\mu$ is the magnification. This is intended to
simulate a realistic survey in which systems with a maximum-time image
are sought first, and then followed up with more sensitive
observations to detect the additional saddle-point image.

In Figure~2 we plot the fractional area within the radial caustic that
produces detectable central images, as a function of
$M_{BH}$.\footnote{The results can be generalized to any Einstein
radius, black hole mass, cosmology and redshifts by noting that they
depend only on the ratio $M_{BH}/M_{Ein}$, where $M_{Ein}$ is the
projected mass within the Einstein radius.} For each choice of
$\beta$, detectable central images form over a range of $M_{BH}$
spanning less than one decade. The high-mass cutoff occurs where the
SMBH destroys the central images, a threshold that is independent of
the detectability criteria. The low-mass cutoff, in contrast, is
enforced by the lower bounds on $\mu_C/\mu_A$ and $\mu_D/\mu_C$. Thus,
smaller black holes could be detected with more sensitive
observations.

The mass range that produces central image pairs is (fortuitously)
coincident with astrophysically realistic SMBHs. Using an isothermal
approximation to relate the image separation to the velocity
dispersion, we find that the local $M_{BH}$--$\sigma$ relation
(Tremaine et al.\ 2002) predicts that a SMBH of mass $\log
(M_{BH}/M_{\odot}) = 8.2 \pm 0.2$ will reside in our example galaxy --
well within the detectable range for most mass profiles (Fig.~2).

\section{Monte Carlo Simulations}

Next we turn to the question of how accurately the mass of the SMBH
can be recovered from observations of a central image pair. Because a
pair of central images has yet to be detected in any real lens system,
our answer must rely upon simulations.

We create simulated lenses using a PL model with an Einstein radius of
$0\farcs75$, a central point mass, and an external shear of $10\%$. We
consider profiles spanning the range $1.75 \leq \beta \leq 1.95$,
which is consistent with many observed lens galaxies (see, e.g., Rusin
et al.\ 2003; Winn et al.\ 2003b; Treu \& Koopmans 2004; Rusin \&
Kochanek 2005), and produces detectable central image pairs with
reasonable probability. For each $\beta$ we focus on the range of
$M_{BH}$ that is most efficient at creating such systems (Fig.~2).  We
focus exclusively on ``doubles'' (which have two bright images and two
faint central images) because they generally produce much brighter
central images than ``quads'' (Mao et al.\ 2001, Keeton 2003).  We
perturb the simulated lens data using Gaussian errors of 1~mas for the
image positions (typical of radio observations), 20\% for the fluxes
(typical of systematic errors due to mass substructure or
microlensing), and 20~mas for the lens galaxy position.

We re-fit the simulated lenses using models consisting of a spherical
galaxy, a central point mass, and an external shear field. The models
have 4 degrees of freedom. We simultaneously optimize all
parameters using a standard $\chi^2$ statistic including contributions
from the image positions and magnifications, and lens galaxy
position. For an ensemble of lenses created from a given set of input
parameters, we extract the distribution of best-fit values of $\log
M_{BH}$, which we describe by its mean and standard deviation. For
each individual lens we also calculate the range of acceptable $\log
M_{BH}$ using the $\Delta \chi^2$ method, verifying that these
uncertainties correspond closely to the standard deviation of the
ensemble.

First we re-fit the simulated lenses using the parent PL model. We
find that the input black hole mass can be recovered with an accuracy
of $0.05-0.10$ dex in $M_{BH}$. This excellent accuracy is related to
the very tight simultaneous constraint that we derive on the
logarithmic density slope $\beta$. As modeling of the three-image lens
PMN~J1632$-$0033 has shown, the properties of image C are
sufficient to constrain the PL slope (Winn et al.\ 2003b). The
properties of image D then determine $M_{BH}$.

The preceding results assume that the functional form of the galaxy
mass distribution is known. In reality, uncertainty in the central
mass profile can lead to a systematic error. To investigate this
error, we re-fit the simulated lenses with a NIS profile instead of
a PL profile. These two models are qualitatively different in the
inner region where central images form, yet the lensing data alone may
not be able to discriminate between them (Winn et al.\ 2003b). We find
that a pair of central images does no better than a single central
image in distinguishing between PL and NIS galaxies. Furthermore,
the NIS model yields black hole masses that are
systematically larger than the ``true'' (PL) values. Figure~3 shows
the systematic offset at a single input $M_{BH}$ for different input
$\beta$. The offset is $0.6$ dex for $\beta = 1.75$, and increases to
$1.0$ dex at $\beta = 1.95$.\footnote{The offset rises very rapidly as
$\beta\rightarrow 2$. For example, the systematic error is $1.4$ dex
at $\beta = 1.98$, up from $1.0$ dex at $\beta=1.95$. This is of
little concern, however, as the cross section for producing detectable
central image pairs is very small for $\beta > 1.95$.} For the PL
slope that best fits PMN~J1632$-$0033 ($\beta=1.90$), the offset is
$0.8$ dex.

We believe that the PL versus NIS analysis gives an upper bound on the
systematic error in $M_{BH}$, because there is a finite range of inner
slopes for a realistic galaxy, and the PL and NIS models bracket this
range. First, the NIS model has the shallowest possible inner profile
(a finite-density core) because the density is expected to decrease
monotonically with radius. Second, galaxy mass profiles are always
observed to become shallower at smaller radius, rather than steeper
(Byun et al.\ 1996; Ravindranath et al.\ 2001; Trujillo et al.\
2004). Hence the inner profile is not steeper than the best-fit global
power-law profile.

To show explicitly that the systematic error depends mainly upon the
inner logarithmic slope of the mass distribution, we next consider a
broken power-law (BPL) model, for which the surface density varies as
$R^{-\gamma_{in}}$ for $R < R_0$, and as $R^{-\gamma_{out}}$ for $R \geq
R_0$. By fixing the outer slope at the isothermal value ($\gamma_{out}
= 1$), the BPL can approximate both the NIS (for $\gamma_{in} = 0$)
and the PL (for $R_0 \rightarrow \infty$, $\gamma_{in} = \beta
-1$). In Figure~4 we show the results of fitting the BPL model to
simulated lenses generated with a PL model.

The BPL analysis yields three main results. First, as before, we find
that it is not possible to distinguish among different values of
$\gamma_{in}$. Second, we find that the recovered $\log M_{BH}$ varies
systematically with $\gamma_{in}$, bridging the PL and NIS
limits. Third, we find that the inferred $\log M_{BH}$ varies less
strongly with $\gamma_{in}$ when that slope is shallow, as opposed to
steep. Most of the variation in $\log M_{BH}$ occurs just as the model
is reaching the PL limit. Thus, even a crude determination of
$\gamma_{in}$ from other observations should substantially reduce the
systematic error. Empirically, Byun et al.\ (1996) and Ravindranath et
al.\ (2001) found that the majority of nearby ellipticals have inner
luminosity profiles with $\gamma_{in} < 0.6$.

Can additional constraints help to differentiate among galaxy models?
We investigated three ideas:\ (1) a high-precision (1~mas) measurement
of the galaxy center, as might be available if the SMBH acts as an
AGN; (2) well-measured time delays ($\sim\!0.5$ day) among all images;
and (3) a requirement that the mass quadrupole be aligned with the
observed position angle of the lens galaxy to within $20^{\circ}$
(see, e.g., Kochanek 2002; Winn et al.\ 2003a). Unfortunately, our
simulations show that these constraints, even in combination, do not
break the degeneracy between the inner density profile and
$M_{BH}$. The time delays among B, C and D are typically too short to
be useful, and the shear constraint adds little information. Most
promising is the constraint on the black hole position, which in some
cases can distinguish between the PL and NIS models at the
2--3~$\sigma$ level. Of course, it is not necessarily true that the AGN
precisely marks the center of the galaxy mass distribution.

\section{Summary and Discussion}

We have investigated the power of a central image pair to measure the
mass of a SMBH. By creating and modeling simulated lens systems, we
draw two main conclusions. First, if we know the form of the galaxy
mass distribution over the range of radii where central images
appear, then realistic observations will determine $M_{BH}$ to within
$0.1$ dex. Second, uncertainty in the inner mass distribution
introduces a significant systematic error in $M_{BH}$. Our simulated
lens data were fitted equally well with a power law (PL) and a
nonsingular isothermal sphere (NIS), but the inferred values of
$M_{BH}$ were 0.6--1.0~dex larger for the NIS fit. On physical and
empirical grounds, we believe that this is the maximum systematic
error in $M_{BH}$. We also explicitly demonstrated the dependence of
$M_{BH}$ on the inner profile slope, using a broken power-law (BPL)
model.

Despite the systematic error, we believe that the lensing technique
will be useful. We are not aware of any other proposed method for the
direct measurement of black hole masses in distant, inactive galaxies.
Furthermore, we note that the maximum systematic error for SMBHs at
{\em intermediate redshift} is only 2--3 times larger than the
statistical error in {\em local} black hole measurements. The
systematic error could be eliminated by directly measuring the inner
luminosity (and hence mass) profile of the lens galaxy, although this
will be very difficult. An alternative would be to apply priors based
on observations of local ellipticals.

Apart from the high sensitivity needed to detect central image pairs,
there are additional observational challenges. First, one must rely on
the usual lens tests -- a common spectrum, surface brightness, and
correlated variability -- to show that a radio source is actually a
central image rather than an AGN in the lens galaxy. Second,
high-frequency observations may be required to minimize the effects of
free-free absorption (Winn et al.\ 2003b). Finally, stellar
microlensing and halo substructure can perturb the magnifications of
central images, complicating the interpretation of measured flux
densities. Such effects, however, are not expected to significantly
degrade the accuracy of the method, mainly because of the large
angular size of the background radio source (G. Dobler et al., in
preparation).

Future observations with the Square Kilometer Array offer the best
hope for overcoming these issues and discovering large samples of
lenses with central image pairs. Since the outer images reveal the
galaxy mass, and the central image pairs determine the black hole
mass, this instrument may allow us to investigate the evolution of the
black hole--galaxy connection.

\acknowledgements

We thank Avi Loeb, Chris Kochanek, and Greg Dobler for helpful
discussions. Work by J.N.W. was supported by NASA through Hubble
Fellowship grant HST-HF-01180.02-A, awarded by the Space Telescope
Science Institute, which is operated by the Association of
Universities for Research in Astronomy, Inc., for NASA, under contract
NAS~5-26555.

\clearpage

\begin{figure*}
\psfig{file=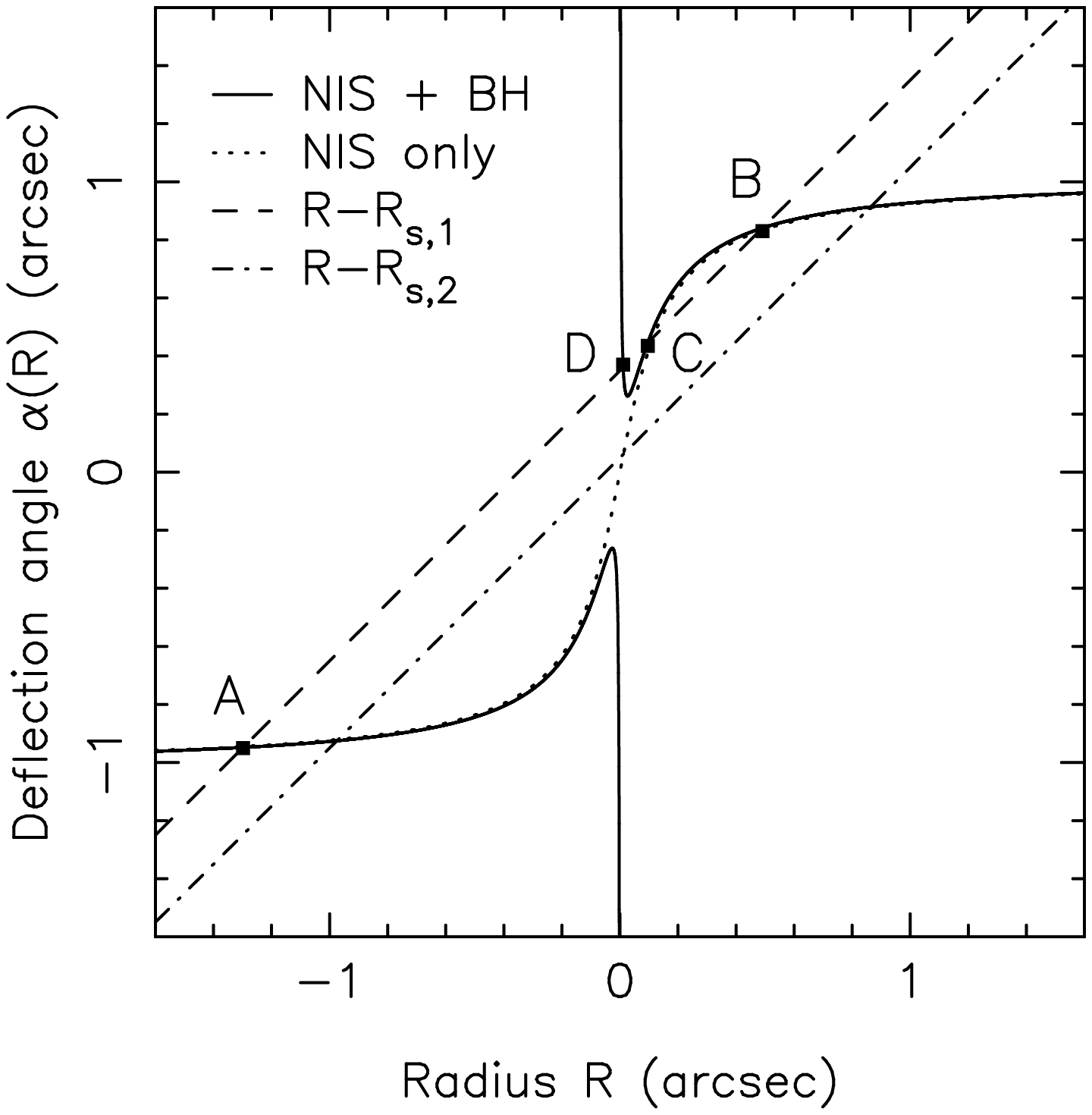,width=3.5in}
\figurenum{1}
\caption{Phenomenology of central images, as illustrated by a NIS
model. We plot the deflection profile $\alpha(R)$ with (solid line)
and without (dotted line) a central black hole. Intersections of
$\alpha(R)$ and the line $R-R_s$ (where $R_s$ is the source-plane radius)
mark the locations of lensed images. The NIS alone produces three
images (A, B, C) of any source within the radial caustic. In the
presence of a central black hole, a fourth image (D) is produced for
sufficiently misaligned sources ($R_{s,1}$: dashed line), while for
well-aligned sources image C is destroyed ($R_{s,2}$: dash-dotted
line).}
\end{figure*}

\clearpage

\begin{figure*} 
\psfig{file=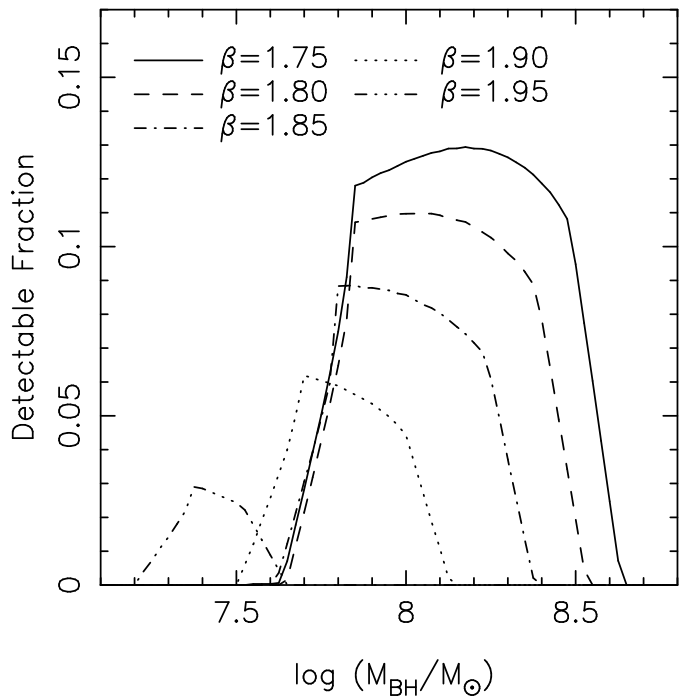,width=3.5in} 
\figurenum{2} 
\caption{Detectability of central image pairs produced by a PL
galaxy with a supermassive black hole. We plot the fraction of source
positions inside the radial caustic that satisfy image magnification
ratio cuts of $\mu_C / \mu_A > 2\times 10^{-3}$ and $\mu_D / \mu_C >
10^{-2}$. The lens redshift is 0.5, the source redshift is 1.5, and
the Einstein radius is $0\farcs75$ for these and all subsequent
simulations. We show results for five different profile slopes,
spanning the range $1.75 \leq \beta \leq 1.95$. According to the local
$M_{BH}$--$\sigma$ relation, the model galaxy should host a SMBH
with $\log(M_{BH}/M_{\odot}) = 8.2 \pm 0.2$.}
\end{figure*}

\clearpage

\begin{figure*}
\psfig{file=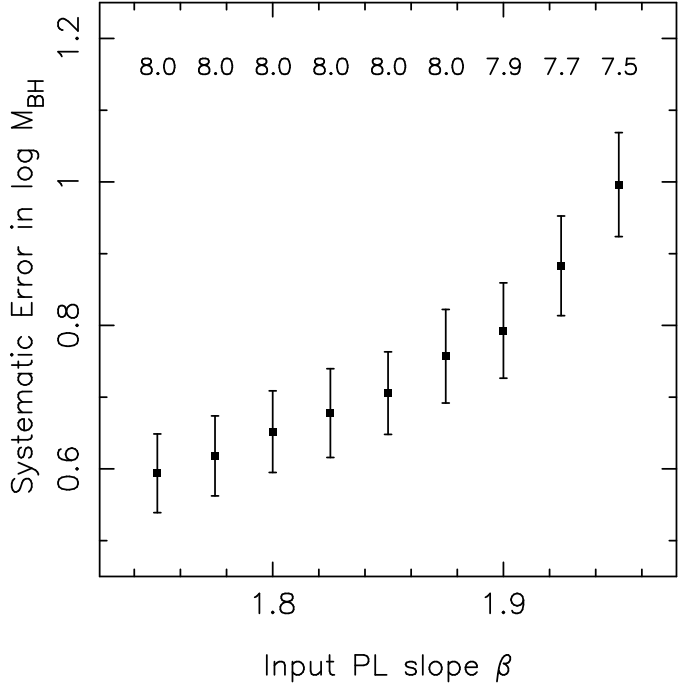,width=3.5in}
\figurenum{3}
\caption{Systematic errors in black hole mass measurement. For each
input PL slope $\beta$, simulated lenses are created using a
representative black hole mass [note the values of
$\log(M_{BH}/M_{\odot})$ listed above the data points] that produces
detectable central images. These lenses are then fitted with a NIS
model. We plot the difference between the mean recovered value of
$\log M_{BH}$ from the NIS model, and the input $\log M_{BH}$. Values
of $M_{BH}$ extracted using the NIS are systematically larger than the
PL values by 0.6--1.0 dex. The dependence of the systematic offset
on $M_{BH}$ at fixed $\beta$ is not plotted, but it is a weak
function, varying by less than 0.1 dex over the entire range of
$M_{BH}$ that produces detectable central images. Moreover, the
systematic offset is not sensitive to the magnification cuts applied
to the simulated lens sample.}
\end{figure*}

\clearpage

\begin{figure*}
\psfig{file=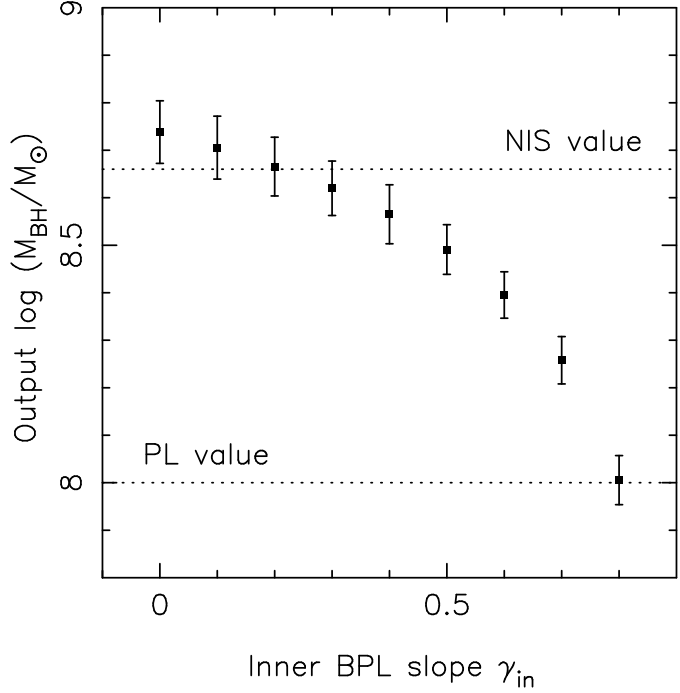,width=3.5in}
\figurenum{4}
\caption{Systematic errors related to the inner profile
slope. Simulated lens systems, created from a PL mass distribution
with slope $\beta = 1.80$ and $\log(M_{BH}/M_{\odot}) = 8.0$
(representative choices), are fitted with a BPL surface density model
with outer slope $\gamma_{out} = 1$. We plot the mean value of $\log
M_{BH}$ extracted from the BPL fit, and find that it varies
systematically with the inner BPL slope $\gamma_{in}$. Also shown are
the input $\log M_{BH}$, and the mean value extracted from a fit using
a NIS, which can be approximated by a BPL with $\gamma_{in} = 0$.}
\end{figure*}

\end{document}